\documentclass{PoS}

\usepackage{amsmath}
\usepackage{amssymb}
\usepackage{longtable}
\usepackage[numbers,sort&compress]{natbib}
\usepackage{url}
\usepackage{wrapfig}

\title{Strong Interactions for the LHC}

\ShortTitle{Strong Interactions for the LHC}

\author{\speaker{George T.\ Fleming}\\
        Yale University, P.O.\ BOX 208120, New Haven CT 06520-8120, USA\\
        E-mail: \email{George.Fleming@Yale.edu}}

\abstract{Dynamical electroweak symmetry breaking (DEWSB) has been a viable
option for the completion of the standard model for over thirty years.
Precision electroweak studies indicate that the new strong interactions that
break EW symmetry cannot be a scaled-up copy of QCD.  Building viable models
of DEWSB is difficult without a detailed understanding of such non-QCD gauge
theories which still confine and break chiral symmetry.  We review past
difficulties of studying these theories using lattice methods and describe
recent progress, focusing on the role of approximate infrared conformal
symmetry.}

\FullConference{The XXVI International Symposium on Lattice Field Theory\\
                 July 14-19 2008\\
                 Williamsburg, Virginia, USA}

\begin{document}

\section{\label{sec:introduction}INTRODUCTION}

The Large Hadron Collider (LHC) will soon probe physics at the TeV scale.
Many theoretical models assume that phenomena at LHC energies involve strongly
interacting systems (for reviews see \cite{Amsler:2008zz,Hill:2002ap,
Lane:2002wv}.) Yet, theories of strong dynamics are very difficult to study
from first principles.  As a result, the predictions of many phenomenological
models are affected by a large margin of uncertainty.  Fortunately, recent
advances in lattice gauge theory (LGT), paralleled by huge increases in
available computational power, have made it possible to derive precise
quantitative information for models based on strong dynamics by numerical
simulation techniques.  In the last year, several groups have begun aggressive
calculational programs to expand our understanding of strongly interacting
gauge theories that are hopefully unlike the very familiar quantum
chromodynamics (QCD).

\section{\label{sub:DEWSB}Dynamical electroweak symmetry breaking}

The current standard model of electroweak interactions was first proposed in
its essentially complete form by Weinberg in 1967 \cite{Weinberg:1967tq}.  At
the time, it was unclear whether this model was renormalizable and thus it was
ignored for four years until its renormalizability was proven by 't Hooft
\cite{'tHooft:1971rn}.  Since then it has become the most successful theory of
particle physics, withstanding more than three decades of testing with ever
increasing precision.  Despite the tremendous efforts of experimental particle
physicists worldwide, two open questions remain.  The first question is how to
incorporate the phenomena of neutrino oscillations and will not be discussed
further here.  The second question is what are the nature of the
Nambu-Goldstone bosons (NGBs) which are eaten to give mass to the electroweak
bosons via the Higgs mechanism.  This review will mostly focus on those
mechanisms of electroweak symmetry breaking (EWSB) which involve new strong
interactions and, thus, where LGT will be the preferred calculational tool for
studying the dynamics of such variants of the electroweak theory.

The original proposal, still very much consistent with experimental data, is
where a single Higgs doublet breaks electroweak symmetry and leaves behind a
single, neutral scalar Higgs boson with a mass no more than a few hundred GeV,
which plays a critical role in making the electroweak theory viable at LHC
energies.  But theoretical advances in the 1970's have led to simple
naturality arguments focused on the renormalization of the Higgs mass which
conclude that this theory must break down at energies not more than a few TeV.
To make the breakdown scale higher and therefore clearly beyond the reach of
the LHC would require some level of fine tuning of the minimal standard model.
Rather than accept such careful fine-tuning, most theorists now conclude that
some new physics described by a theory that extends beyond the standard model
-- a BSM theory -- will manifest itself at LHC energies. Over the last three
decades, a large number of candidate BSM theories has been developed.  They
must satisfy many stringent constraints in order to be viable, and this
excludes many possibilities.  Still, based on current experimental results it
is not possible to select among them.

One of the earliest alternatives to the single Higgs doublet mechanism was
proposed by Weinberg as well \cite{Weinberg:1975gm,Weinberg:1979bn} and is
generally referred to as dynamical electroweak symmetry breaking (DEWSB), or
sometimes \textsl{technicolor} \cite{Susskind:1978ms} to emphasize the analogy
with QCD.  In the minimal technicolor model, a new strong force similar to QCD
but operating at the TeV scale (three orders of magnitude larger than the QCD
scale) plays a central role, driving electroweak symmetry breaking.  In this
model a host of new TeV-scale resonances, analogous to the $\rho$, $\pi$,
$\omega$, \textit{etc.}\ of QCD, may be visible at the LHC.  Unfortunately, a
technicolor sector with the same low energy structure as QCD but scaled up to
the electroweak scale is phenomenologically disfavored, as described below.
However, this creates an opportunity for LGT specialists to survey the
low energy landscape of spontaneously broken gauge theories for those with
a sufficient number of NGBs and other low energy structure that may be quite
different than QCD but favored by the phenomenology of DEWSB.

A much studied possibility in recent years is that space may have extra
dimensions which are very small in size and therefore have not been observed
up to now, but could be directly seen at LHC energies.  In some
extra-dimensional scenarios, the effective four-dimensional theory at LHC
energies is weakly coupled, while in others the effective theory exhibits
strong-coupling phenomena almost indistinguishable from technicolor theories.
In fact, this similarity between some extra-dimensional models and technicolor
models, typically referred to as ``holography'' or ``Ads/CFT'', has been
employed by model builders to gain some insight into strong dynamics.  This
creates further opportunities for collaboration with LGT specialists to match
these five-dimensional effective theories with their corresponding
four-dimensional, UV-complete gauge theories.  See review in these proceedings
by E.~Katz \cite{Katz:2008}. 

\section{\label{sec:TC}\textsl{WASN'T TECHNICOLOR RULED OUT A DECADE AGO?}}

\subsection{\label{sub:Sparam}The Peskin-Takeuchi $S$ parameter}

\begin{figure}
  \begin{center}
    \includegraphics[width=\textwidth]{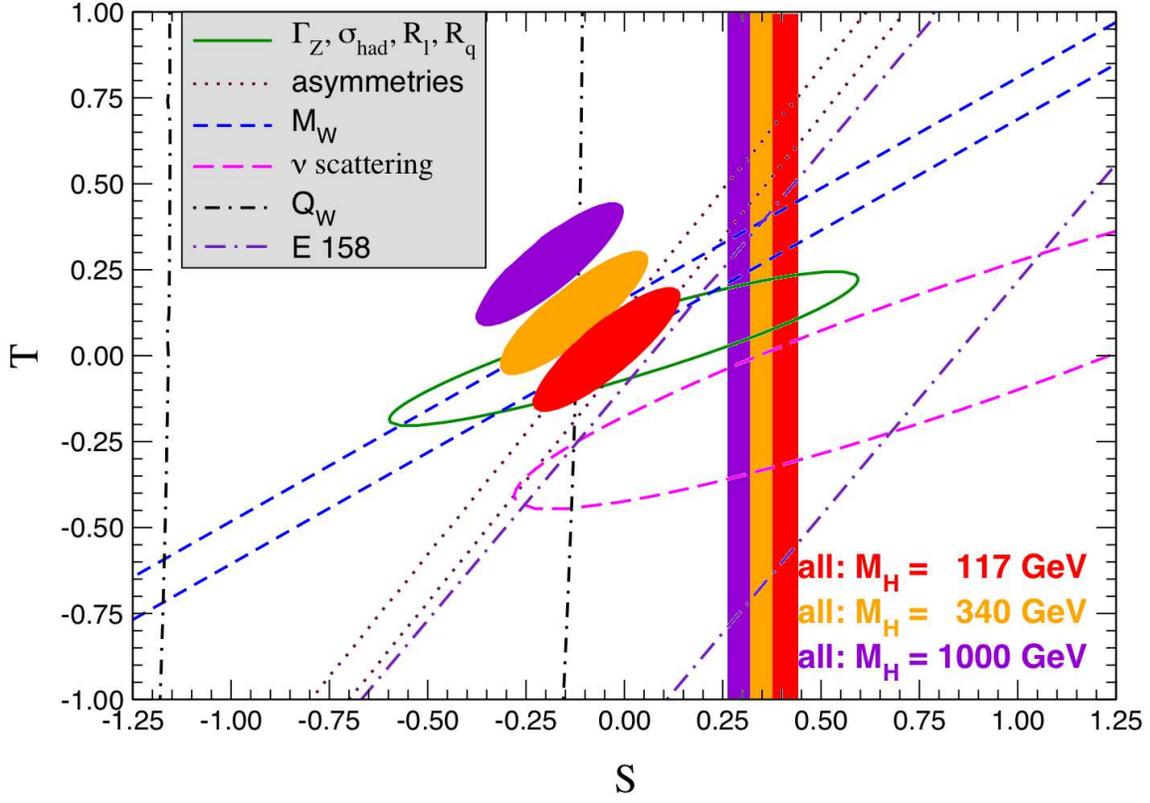}
  \end{center}
  \caption{\label{fig:ST_2007} The shaded ellipses are the phenomenologically
    allowed regions for $S$ and $T$ at 68\% CL, and show a strong correlation
    (87\%) between $S$ and $T$ \cite{Amsler:2008zz,Grunewald:2005ema}.  The
    shaded vertical bands are the recently computed values of $S$ in
    technicolor with $N_\mathrm{TC}=3$ and $N_\mathrm{TF}=2$ at 68\% CL
    \cite{Shintani:2008qe}.  The color of the shaded region indicates the
    reference value for the Higgs mass.}
\end{figure}

In the absence of any direct evidence for new particles that would reveal the
physics of EWSB, three decades of effort starting with Veltman
\cite{Veltman:1976rt,Veltman:1977kh} have gone into using precision
electroweak parameters to limit the number of possible EWSB theories.  A
particularly useful pair of parameters, $S$ and $T$, were devised by Peskin
and Takeuchi \cite{Peskin:1990zt,Peskin:1991sw} to place severe constraints on
models of DEWSB.  These constraints assume that changing the number of flavors
($N_\mathrm{TF}$) or colors in SU($N_\mathrm{TC}$) gauge theories mainly
affects the counting of quark degrees of freedom.  They must further assume
important non-perturbative features remain unaltered: confinement, chiral
symmetry breaking and the ordering of states in the hadronic spectrum.  This
will be called the \textsl{scaled-up QCD} hypothesis in this review.  Assuming
scaled-up QCD, $S$ in technicolor are estimated to be bounded by the very
crude relation:
\begin{equation} \label{eq:S_approx} S \gtrsim \frac{1}{6\pi}\left[
\frac{N_\mathrm{TC} N_\mathrm{TF}}{2} \right] \end{equation}
The latest phenomenological constraints on $S$ and $T$, shown in
Fig.~\ref{fig:ST_2007}, when combined with the naive scaling relation of
Eq.~(\ref{eq:S_approx}) leads the Particle Data group to conclude:
\textit{This rules out simple Technicolor models with many techni-doublets and
QCD-like dynamics} \cite{Amsler:2008zz}.  In other words, technicolor models
with QCD-like dynamics and $N_\mathrm{TC},\ N_\mathrm{TF} \lesssim 4$ are
still consistent with, but somewhat disfavored by, precision electroweak
experiments.  Since the allowed range still includes QCD itself, it is
important to directly calculate $S$ in lattice QCD and compare with estimates
like Eq.~(\ref{eq:S_approx}) to show there are no underestimated systematic
errors in the phenomenological approach.

The basic observable that must be computed in QCD to estimate precision
electroweak parameter $S$ in scaled-up QCD is the low energy part of the
left-right current-current correlation function
\begin{equation}
\label{eq:LR_correlator}
\Pi_{LR}^{\mu\nu}(q) =
i g^{\mu\nu} \Pi_{LR}(q^2) + (q^\mu q^\nu\ \mathrm{terms})
\equiv \int d^4x e^{-iqx} \left\langle J_L^\mu(x) J_R^\nu(0) \right\rangle
\end{equation}
The low energy part is given by the slope of the correlation function
\begin{equation}
\Pi_{LR}^\prime(0) = \left. \frac{d}{dq^2} \Pi_{LR}(q^2) \right|_{q^2 \to 0}.
\end{equation}
If we use the identity $J_{L,R}^\mu(x) = J_V^\mu(x) \pm J_A^\mu(x)$ then the
low energy part is also given by $\Pi_{LR}^\prime(0) = \left[
\Pi_{VV}^\prime(0) - \Pi_{AA}^\prime(0) \right]$.

Once the current-current correlator has been computed, there are still several
steps required to extract the precision electroweak parameters.  In
next-to-leading-order (NLO) chiral perturbation theory
\cite{Appelquist:1980vg,Longhitano:1980iz,Longhitano:1980tm,Gasser:1983yg}
\begin{equation}
\Pi_{LR}^{\mu\nu}(q)
= \left[ g^{\mu\nu} - \frac{q^\mu q^\nu}{m_\pi^2 - q^2} \right] f_\pi^2
  + \left( q^\mu q^\nu - g^{\mu\nu} q^2 \right)
  \left[
    \frac{1}{3}\left( 1 - \frac{4 m_\pi^2}{q^2} \right) \overline{J}(q^2)
    + \frac{1}{48\pi^2} \left( \overline{l}_5 - \frac{1}{3} \right)
  \right]
\end{equation}
where the goal is to determine $\overline{l}_5$.  Note however that the first
term has a pion pole contribution that must be included in any fit to the
lattice results.  The two-body phase space integral $\overline{J}(q^2)$ arises
from the two-pion intermediate state, leading to a cut starting at
$q^2=4m_\pi^2$.

The low energy constant $\overline{l}_5$ is logarithmically divergent in the
chiral limit and must be cut off, introducing some scale dependence
\begin{equation}
l_5^r(\mu) = \overline{l}_5  + \log\frac{m_\pi^2}{\mu^2}
\end{equation}
where $\mu$ is some convenient UV scale, \textit{e.g.}\ in QCD, typically $\mu
= 4 \pi f_\pi$ or $m_\rho$.  In the context of DEWSB, precisely how the
divergence is cut off will depend on the number of NGBs.  Three of the NGBs
will be cut off by the masses of the $W$ and $Z$ bosons when they are eaten.
Extra pseudo Nambu-Goldstone bosons (PNGBs) will be problematic, if present,
unless an additional mechanism, described in Sec.~\ref{sub:FCNC} is provided
to give them masses.   Once the PNGBs are massive, the should behave
analogously to $K$ mesons in the chiral perturbation theory of QCD.   Finally,
the precision electroweak parameters like $S$ are defined to be zero in the
minimal standard model with a single Higgs field.  So, the standard model
contribution must be subtracted, assuming some specific value for the Higgs
mass.  In summary, the low energy constant $l_5^r(\mu)$ computed in QCD is
related to the precision electroweak parameter $S$ by
\begin{equation}
\label{eq:Sparam}
S = \frac{1}{12\pi}
\left[ l_5^r(\mu) + \log \frac{\mu^2}{m_H^2} - \frac{1}{6} \right]
\end{equation}
in the case $N_\mathrm{TF}=2$ with no PNGBs.

The first determination of $\overline{l}_5$ was made recently by the JLQCD
collaboration \cite{Shintani:2008qe}.  The quoted value $L_{10}^r(m_\rho) =
-5.2(^{+7}_{-5})\times 10^{-3}$, converted in to the convention of Peskin and
Takeuchi \cite{Peskin:1991sw} by $l_5^r(\mu) = -192 \pi^2 L_{10}^r(\mu)$, can
be used with Eq.~(\ref{eq:Sparam}) provided $m_\rho$ is rescaled by
identifying $F_\pi$ with the vacuum expectation value of the Higgs field, $v =
2^{-1/4} G_F^{-1/2} = 246\mathrm{\ GeV}$.  Taking $m_\rho = 770\mathrm{\ MeV}$
and the computed value of $F_\pi=87.3(5.6)\mathrm{\ MeV}$ from
\cite{Fukaya:2007pn} yields $S$ = $0.41(^{+3}_{-4})$, $0.36(^{+3}_{-4})$, and
$0.30(^{+3}_{-4})$ for $m_H$ = 117, 340, and 1000 GeV, respectively.  In
Fig.~\ref{fig:ST_2007}, these allowed regions for $S$ are plotted along with
the phenomenologically allowed regions for $S$ and $T$ at 68\% CL.  The
disagreement between the two determinations a bit more than the two sigma
level and essentially independent of the input Higgs mass.

So, the $S$ parameter can be used to place constraints on the number of
technicolors ($N_\mathrm{TC}$) and techniflavors ($N_\mathrm{TF}$) a
technicolor theory may have when the underlying dynamics are like QCD.
However, it is in fact a dynamical question whether a given theory is QCD-like
as $N_\mathrm{TC}$ and $N_\mathrm{TF}$ are varied.  If the dynamics of the
theory are sufficiently different from QCD then the above estimates do not
apply.  In some walking theories (see Sec.~\ref{sub:walking_Yang-Mills} below)
$S$ could be very small and possibly negative.  So, the current
phenomenological bounds on $S$ do not yet place stringent constraints on
technicolor theories due to the lack of understanding how the dynamics of
confining gauge theories change as the number of colors and flavors, or color
representation of the fermions, are varied.

\subsection{\label{sub:FCNC}Fermion masses and flavor changing neutral
currents}

QCD-like technicolor theories with small $N_\mathrm{TC}$ and $N_\mathrm{TF}$
and no extra PNGBs cannot be ruled out alone by precision EW observables like
the $S$ parameter.  However, such a technicolor theory has only explained how
the $W$ and $Z$ bosons have become massive through the Higgs mechanism.
Technicolor alone does not explain how standard model fermions become massive
nor how any additional NGBs beyond the three eaten during DEWSB become massive
PNGBs.  It seems natural to consider additional gauge interactions, called
\textsl{extended technicolor} (ETC) \cite{Dimopoulos:1979es,Eichten:1979ah},
that couple to both standard model fermions and technifermions.  The ETC gauge
symmetry will be broken at some higher scale $\Lambda_\mathrm{ETC}$ and  at EW
energies exchange of massive ETC gauge bosons will be effectively described by
four fermion operators given schematically as
\begin{equation}
  \frac{\left(\overline{Q}Q\right)
  \left(\overline{Q}Q\right)}{\Lambda_\mathrm{ETC}^2} ,
  \qquad
  \frac{\left(\overline{q}q\right)
  \left(\overline{Q}Q\right)}{\Lambda_\mathrm{ETC}^2} ,
  \qquad
  \frac{\left(\overline{q}q\right)
  \left(\overline{q}q\right)}{\Lambda_\mathrm{ETC}^2} .
\end{equation}
where $Q$'s are technifermions and $q$'s are standard model fermions. When
technicolor grows strong enough to break chiral symmetry and form
$\left\langle \overline{Q} Q \right\rangle$ condensates, the first operator
will give mass to uneaten PNGBs and the second operator will give mass to
standard model fermions.  However, the third operator is problematic because
it gives rise to flavor changing neutral currents.  In particular,
experimental measurements of $K^0$--$\overline{K}^0$ mixing and $K_L
\to \mu^+ \mu^-$ as well as upper limits on other rare processes require
$\Lambda_\mathrm{ETC} \simeq 1000 \mathrm{\ GeV}$ \cite{Appelquist:2003hn}.

Since ETC interactions will be highly suppressed at EW energies due to such a
high ETC scale, it is natural to wonder whether standard model fermion masses
can still be generated by this mechanism.  For example, the mass of the
strange quark will be given by $m_s \sim
\left\langle\overline{Q}Q\right\rangle / \Lambda_\mathrm{ETC}^2$.  In QCD, the
chiral condensate  is $\left\langle\overline{q}q\right\rangle / f_\pi^3
\approx 25$.  If $\left\langle\overline{Q}Q\right\rangle / v^3 \sim 25$ as
well, where $v = 2^{-1/4} G_F^{-1/2} = 246\mathrm{\ GeV}$, this leads to very
light strange quarks: $m_s \sim 0.4 \mathrm{\ MeV}$.  Thus, some new dynamics
is required to produce modestly larger techniquark condensates than would
naturally appear in a QCD-like theory.  Walking is one such dynamical
mechanism, described in Sec.~\ref{sub:walking_Yang-Mills}.  Even assuming a
mechanism for enhancing condensates can be realized, flavor changing neutral
currents still pose significant challenges for detailed models of DEWSB
\cite{Appelquist:2003hn}.

\section{\label{sec:flavor_dependence}FLAVOR DEPENDENCE OF SU($N$) YANG-MILLS
  GAUGE THEORY}

In Sec.~\ref{sub:conformal_window}, we review what is known about the flavor
dependence of SU($N$) Yang-Mills theory from the perturbative expansion of the
beta function.  We also mention other non-perturbative and quasi-perturbative
techniques for modeling flavor dependence.  We focus on the question of
whether theory has an infrared fixed point (IRFP) for a given number of
flavors $N_f$.  In Sec.~\ref{sub:finite_temperature}, we review two decades
of efforts to study this problem by determining the flavor dependence of the
finite temperature phase transition. In Sec.~\ref{sub:running_coupling}, we
review past and current efforts to understand the low energy dynamics of
Yang-Mills by studying the non-perturbative running of the renormalized
coupling.

\subsection{\label{sub:conformal_window}Yang-Mills Conformal Window}

The renormalization group describes the scale dependence of the renormalized
coupling of Yang-Mills in a general scheme:
\begin{equation}
L \frac{\partial}{\partial L} g(L) = \beta(g) \overset{g \to 0}{\sim}
b_0 g^3 + b_1 g^5 + b_2 g^7 + \cdots
\end{equation}
In the asymptotic expansion ($g \to 0)$ the first two universal coefficients
are independent of the scheme:
\begin{equation}
b_0 = - \frac{1}{(4 \pi)^2} \left( \frac{11}{3} N_c - \frac{2}{3} N_f \right),
\qquad b_1 = - \frac{1}{(4 \pi)^4} \left[
  \frac{34}{3} N_c^2
  - \left( \frac{13}{3} N_c - \frac{1}{N_c} \right) N_f
\right].
\end{equation}
Of course, $b_0 < 0$ for $N_f < \frac{11}{2} N_c$ implies that SU($N_c$)
Yang-Mills gauge theory with relatively few number of flavors is
\textsl{asymptotically free}.

\begin{figure}[ht]
  \begin{center}
    \includegraphics[width=0.5\textwidth]{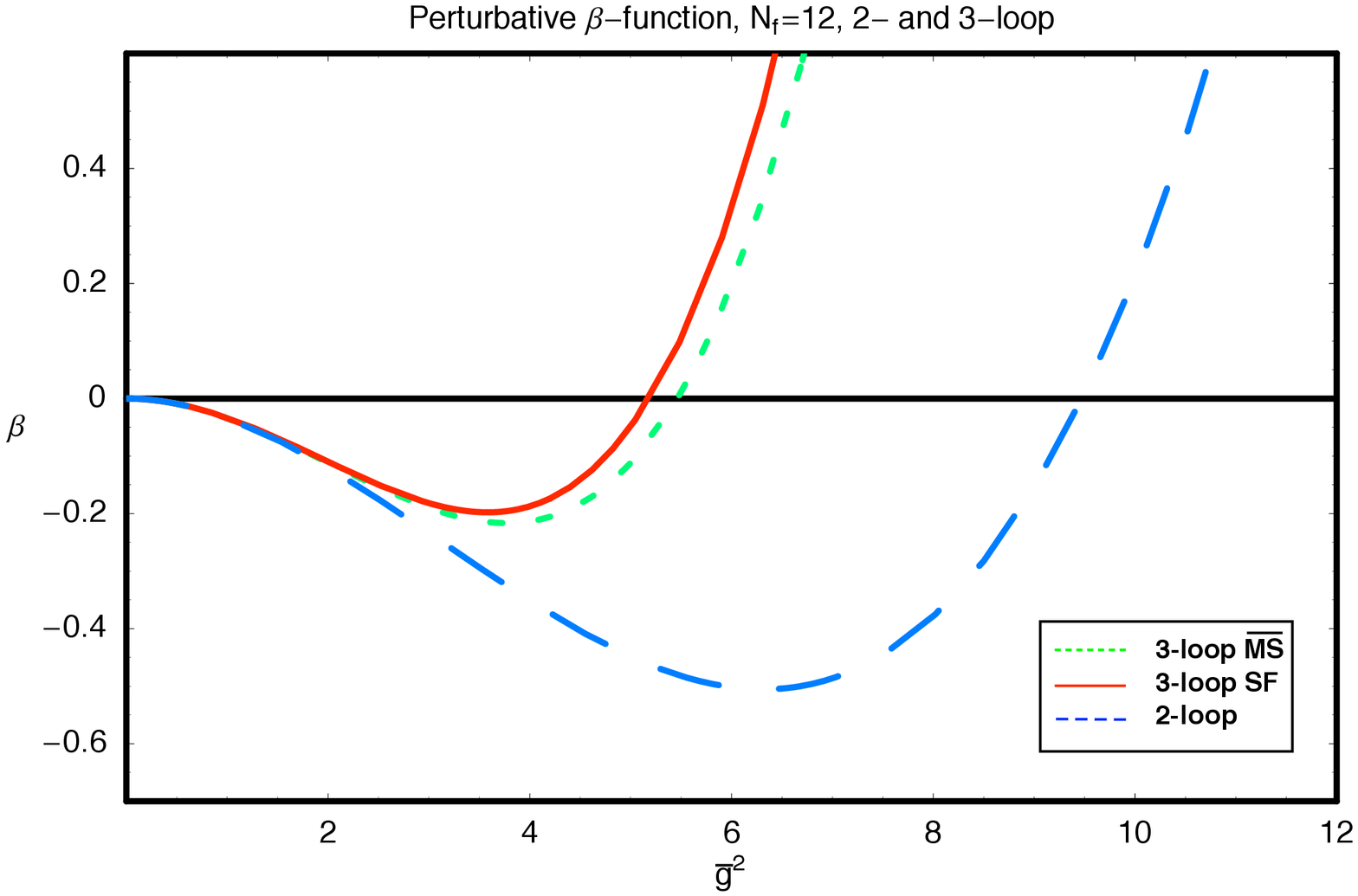}\includegraphics[width=0.5\textwidth]{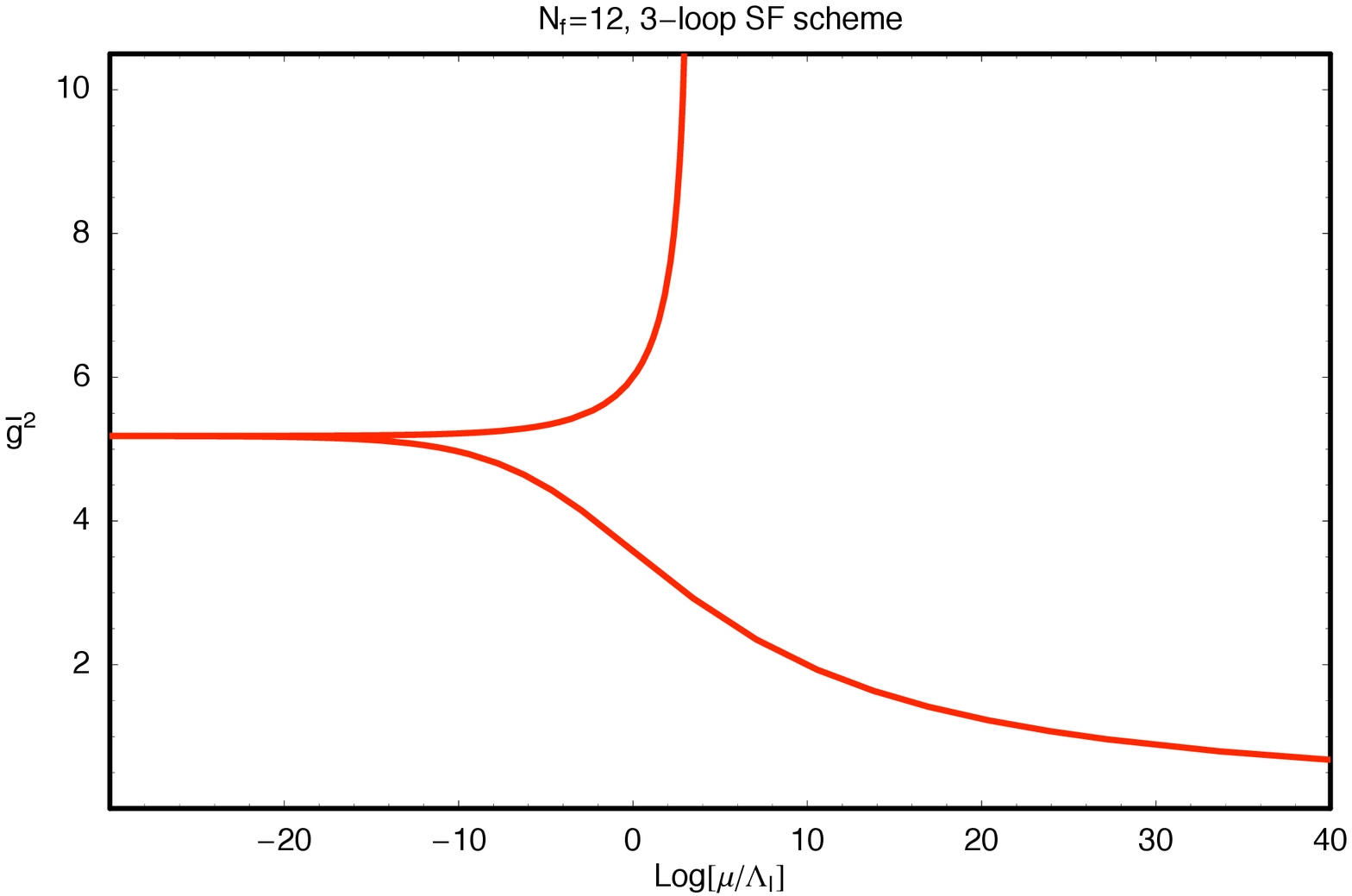}
  \end{center}
  \vspace{-2ex}
  \caption{
    \label{fig:SF_pt_running}
    The left panel shows the two and three loop beta function in the
    Schr\"odinger Functional scheme with $N_f$=12 flavors \cite{Bode:1999sm}.
    The right panel shows the running of the coupling to the IR fixed point
    from both sides.  The reference scale $\Lambda$ may be different on each
    branch.
  }
\end{figure}

For $N_f \lesssim \frac{11}{2} N_c$, the theory is still asymptotically free
but the two loop term $b_1$ can cancel $b_0$ forming a perturbative infrared
(IR) fixed point \cite{Gross:1973ju,Caswell:1974gg,Banks:1981nn}.  As $N_f$
steadily decreases from $\frac{11}{2} N_c$, the magnitude of the IR fixed
point increases until higher order perturbative and even non-perturbative
effects become significant, at which point even the definition of the
renormalized coupling becomes scheme dependent.  Regardless of the choice of
scheme, the physics of the IR fixed point is made clear by the approximate
conformal behavior of the theory at long distances.  Such theories are said to
be in the \textsl{conformal window}.  Eventually, when $N_f$ is sufficiently
small, the conformal window closes when the gauge interactions become strong
enough to confine and/or spontaneously break chiral symmetry.

\subsection{\label{sub:finite_temperature}Flavor Dependence of Finite
  Temperature Transition}

It would seem that lattice field theory techniques are ideally suited to
determine the size of the conformal window in Yang-Mills.  Yet, several
efforts in the 1980's
\cite{Kogut:1985pp,Gavai:1985wi,Attig:1987mf,Kogut:1987ai,Meyer:1990xd} and
again in the 1990's
\cite{Kim:1992pk,Brown:1992fz,Damgaard:1997ut,Iwasaki:2003de} failed to
produce a consistent picture of the zero temperature transition from a few
flavors of confined quarks to many flavors of deconfined quarks in the
non-Abelian Coulomb phase of the conformal window.  This is in sharp contrast
to the incredibly successful use of the lattice regulator to study the finite
temperature transition of QCD during precisely the same period.

\begin{figure}
  \begin{center}
    \includegraphics[height=0.3\textheight]{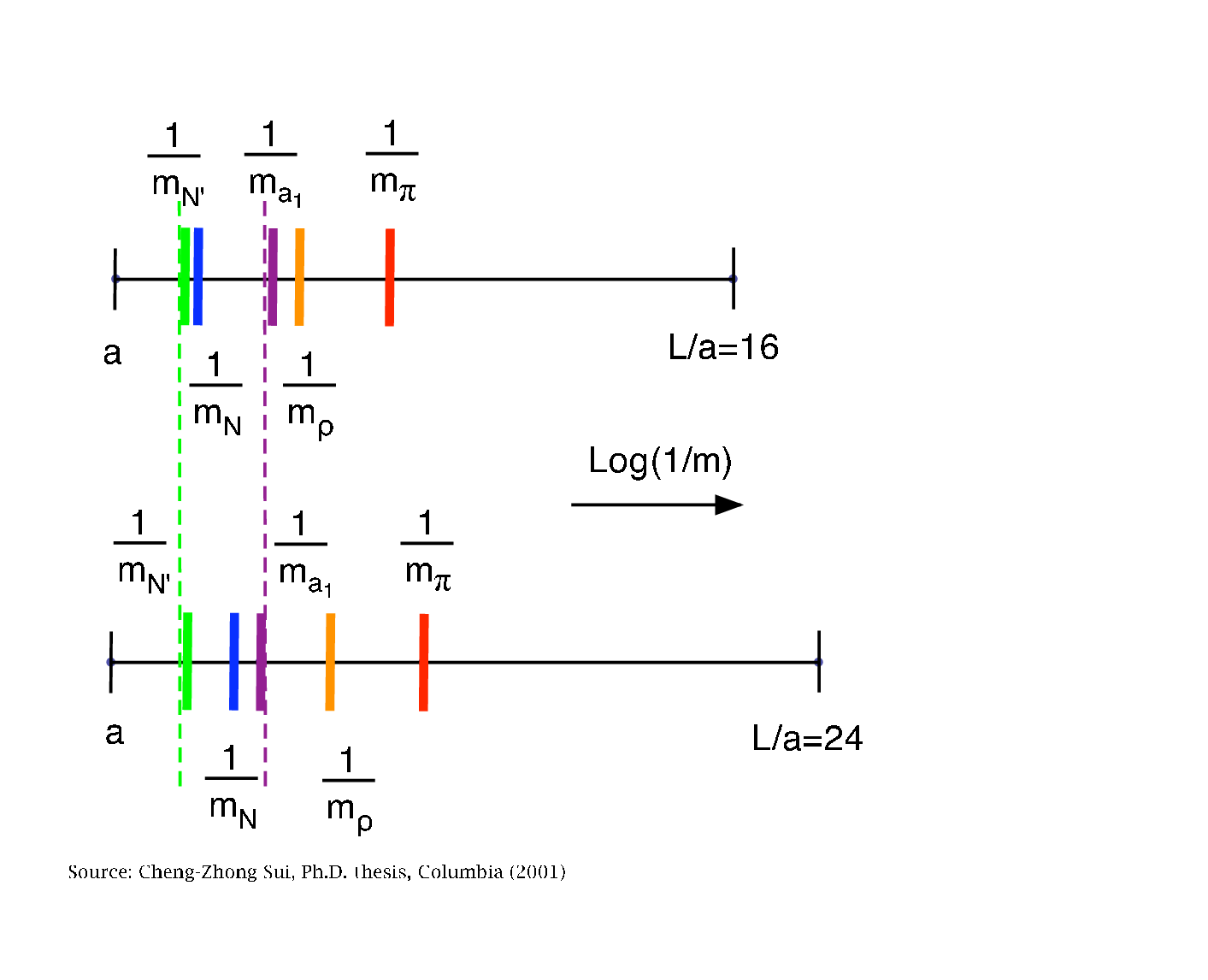}
  \end{center}
  \caption{\label{fig:columbia4f-plotspec-Nprime}Finite volume effects at
    fixed $\beta$ (thus fixed $a$) for $N_f=4$ staggered fermions
    disproportionally affect $\rho$ meson and $N$ nucleon and not their parity
    partners \cite{Sui:2001rf}.}
\end{figure}

In Appendix \ref{app:finite_temperature}, we have summarized the relevant
calculations from this period for staggered fermions with the unimproved SU(3)
Wilson gauge action, with particular focus on $N_f \ge 3$ degenerate fermions.
Although some studies exist with Wilson fermions \cite{Iwasaki:2003de} and
staggered fermions with various forms of improvement, this forms the bulk of
all such studies to determine the size of the conformal window.  The general
trend of the data is clear:  as the number of flavors increases, the critical
temperature decreases relative to some dimensionful quantity computed at
zero temperature, say $T_c / m_\rho$.
Although each of the previous decades' efforts were encouraged by the
theoretical, algorithmic and computational advances of the time, the failure
to make progress appears related to the method of choice for defining the
conformal window. In principle, one could study the finite temperature phase
transition as the number of flavors increases and extrapolate this line of
transitions to the critical number of flavors where the temperature vanishes.
In practice, this proved to be quite challenging due to various difficulties
in making the quarks light enough or the spatial box big enough
(see Fig.~\ref{fig:columbia4f-plotspec-Nprime}).

Recently, two new groups have revived this approach. The Groningen-Frascati
group is focused on eight and twelve flavors of improved (\texttt{asqtad})
staggered fermions \cite{Deuzeman:2008sc,Deuzeman:2008pf,Deuzeman:2008da}.
The Columbia group is focused on eight flavors of unimproved staggered
fermions with the DBW2 gauge action \cite{Jin:2008}.  Both groups provide new
evidence that SU(3) Yang-Mills with eight flavors is a confining theory with
spontaneously broken chiral symmetry and a lattice spacing that vanishes as
the bare coupling approaches zero.  Apparently, the use of improved actions as
well as more computational power was all that was needed to resolve some of
the questions raised by the earlier work of the Columbia group at eight
flavors \cite{Brown:1992fz}.  Both groups are interested in applying their
methods to study SU(3) Yang-Mills with twelve flavors where it is likely the
theory is governed by an IRFP at low energies.

\subsection{\label{sub:running_coupling}Flavor Dependence of the Running
  Coupling}

In hindsight, the calculations of Damgaard \textit{et
al}.~\cite{Damgaard:1997ut} and Heller \cite{Heller:1997pn,Heller:1997vh} of
Yang-Mills with $N_c=3$ colors and $N_f=16$ flavors revealed the basic problem
with the approach: in order to see chiral symmetry breaking on a finite
temperature lattice with relatively few lattice sites in the temporal
direction the bare lattice coupling must be made relatively strong.  Using the
renormalization group, this implies that the lattice spacing is relatively
large and the temperature is relatively low.  However, if the dynamics at the
scale of the lattice spacing are strong enough, using the renormalization
group to take the continuum limit may not be possible.  In the 16 flavor case,
the IR fixed point is completely understood in perturbation theory and no
chiral symmetry breaking should occur.  Yet, Damgaard \textit{et al}.\
observed such a transition using the same techniques employed by other groups
with smaller numbers of flavors.  They surmised, and Heller later showed, that
the chiral transition occurred along a renormalization group trajectory where
the gauge dynamics becomes arbitrarily strong at short distances.  In other
words, there is no continuum limit for the apparently observed finite
temperature transition because it occurs on the wrong side of the IR fixed
point.  This is shown in Fig.~\ref{fig:SF_pt_running} for twelve flavors.
Sixteen flavors is similar except the fixed point value for the running
coupling is much smaller.  The point is that the continuum limit doesn't exist
if the coupling flows to the IR fixed point from above because the theory
flows to strong coupling in the ultraviolet.  Thus, it is reasonable to
question the validity of the other aforementioned calculations since it is not
at all clear that they are connected to the asymptotically free continuum
limit by the renormalization group.

There are in fact several reasonable, non-perturbative definitions of the
Yang-Mills running coupling which all coincide to two loops in perturbation
theory.  One example can be derived from the static quark potential $V(R)$.
For an infinitely heavy quark-antiquark pair separated by a distance $R$ the
force $F(R)$ felt by the pair is defined to be proportional to the
renormalized coupling at the distance $R$
\begin{equation}
\alpha_{Q\overline{Q}}(R) = \frac{g^2_{Q\overline{Q}}(R)}{4\pi} \equiv
\frac{1}{C_F} R^2 F(R), \qquad F(R) = \frac{\partial}{\partial R} V(R)
\end{equation}
For pure Yang-Mills ($N_f=0$), the confining potential grows linearly at long
distances so $\alpha_{Q\overline{Q}}(R) \overset{R\to\infty}{\sim} R^2$.
For a spontaneously broken theory ($N_f > 0$), the color flux tube responsible
for the linear rise in the potential will eventually break when the
static-light meson-antimeson state becomes energetically favored.  Without
some residual attractive force between the two static-light mesons, the 
renormalized coupling $\alpha_{Q\overline{Q}}(R)$ will rapidly vanish.
In the conformal window, the force between static quarks is given by
a non-Abelian Coulomb potential with the long distance coupling
set by the IR fixed point.

In principle, the static quark potential is an ideal tool to search for the
edge of the conformal window.  In practice, the computational costs are such
that the running of $\alpha_{Q\overline{Q}}(R)$ has only been computed out to
confining distances in the quenched theory \cite{Necco:2001xg,Necco:2001gh}.
In a typical zero temperature lattice simulation with light dynamical quarks
in a periodic box, there are four scales in the problem:  the box size
$L$, the light quark mass $m^{-1}$, the potential scale $R$ and the
lattice spacing $a$.  In general, all four scales should be widely separated,
suggesting that $L > 64 a$ at least, which is currently impossible given the
available computational resources and the performance of the best available
algorithms.

To make the calculation of the running coupling tractable in a practical
lattice simulation given available computational resources, the coupling
should be computed on a scale of $O(L)$ to eliminate the required scale
separation $L \gg R$.  A new non-perturbative scheme was presented at this
meeting by M.~Kurachi and E.~Itou \cite{Bilgici:2008mt} where square Creutz
ratios \cite{Creutz:1980wj} calculated at a fixed fraction of the spatial box
size can be used to define a renormalized coupling.  This eliminates one of
the separate scales needed in the static potential scheme and is therefore
much more likely to be tractable in calculations with dynamical fermions.
Further work is underway to eliminate the need for a non-zero quark mass using
twisted boundary conditions.

Another non-perturbative definition of the running coupling is to perform
calculations in a constant background chromoelectric field.  The Schr\"odinger
functional (SF) is the partition function describing the quantum mechanical
evolution of a system from a prescribed state at time $t = 0$ to another state
at time $t = T$ in a spatial box of size $L$ with periodic boundary conditions
\cite{Luscher:1992an,Sint:1993un,Bode:2001jv}.  Dirichlet boundary conditions
are imposed at $t = 0$ and $t = T$ where $T$ is $O(L)$.  They are chosen such
that the minimum-action configuration is a constant chromo-electric background
field of strength $O(1/L)$. This can be implemented in the continuum
\cite{Luscher:1992an} or with lattice regularization \cite{Bode:1999sm}.  In
either case, by considering the response of the system to small changes in the
background field, a gauge invariant running coupling can be defined, valid for
any coupling strength.

Heller had the original idea of	studying the running of the SF renormalized
coupling to look for flow towards an IR fixed point
\cite{Heller:1997pn,Heller:1997vh}.  With very little computational effort, he
clearly demonstrated that the $N_f=16$ chiral phase transition observed by
Damgaard \textit{et al}.~\cite{Damgaard:1997ut} was triggered by strong
dynamics at the UV scale and at long distances the coupling was flowing
rapidly in the general direction of the very weak IR fixed point of $N_f=16$
flavors \textsl{from above}.  Of course, to compute non-perturbatively the
numerical value of such a weak fixed point would have required a great deal of
computing power to statistically resolve it from zero.

In the conformal window, the coupling at the IR fixed point $g^*$ increases as
the number of flavors decreases.  So, for smaller $N_f$, while it may be
desirable to continue to work on the ``wrong'' side of the IR fixed point due
to the relatively rapid flow towards the fixed point there may be very little
``room'' in which to work without triggering a bulk lattice phase transition.
On the asymptotically free side of the IR fixed point, the flow may be gradual
and difficult to distinguish from lattice artifacts due to finite lattice
spacing, but there will be confidence that an extrapolation to the continuum
limit is still possible.

Deeper into the conformal window of SU($N_c$) gauge theories, it is possible
the IR fixed point $g^*$ may become sufficiently large that it is no longer
well described by perturbation theory yet not strong enough to trigger
spontaneous symmetry breaking.   In SUSY QCD, this scenario leads to a
non-trivial duality between the perturbative description defined in
asymptotically free limit and a low energy description of new degrees of
freedom that emerge at IR scales (see \cite{Intriligator:1995au} for a
review).  There are arguments to suggest the IR fixed point of
non-supersymmetric SU($N_c$) gauge theories will remain perturbative
throughout the conformal window \cite{Gardi:1998ch}.  So, a lattice
calculation that could determine $g^*$ with sufficient accuracy to show
disagreement with perturbation theory raises the possibility that something as
interesting as Seiberg duality may occur near such a non-perturbative IR fixed
point.

Recently, we presented the first non-perturbative evidence for the existence
of an IRFP in SU(3) Yang-Mills with twelve flavors \cite{Appelquist:2007hu}.
Additional evidence was presented that SU(3) with eight flavors shows no signs
of an IRFP.  Further details of the calculation were presented by Neil
\cite{Neil:2008} at this meeting.  The main observable computed was the
running coupling in the SF scheme.  Questions as to the scheme dependence of
the conclusion are unavoidable and, thus, it is crucial to test this
conclusion by other means.  The work of the Columbia group and the
Groningen-Frascati group, already mentioned in
Sec.~\ref{sub:finite_temperature}, are important cross checks because they do
rely on observables related to confinement and chiral symmetry breaking.  We
hope that the efforts of Kurachi, Itou and collaborators will soon result in a
calculation of the running coupling in their scheme for twelve flavors to test
the scheme-independence of the IRFP.

Finally, a calculation of the distribution of low-lying eigenvalues of the
staggered Dirac operator was presented by Holland \cite{Fodor:2008hn}.  In
this work, the authors present preliminary evidence that eigenvalue
distributions for eight and twelve flavors of staggered fermions follow the
predictions of random matrix theory for the epsilon regime ($M_\pi \ll L^{-1}
\ll F_\pi \ll a^{-1}$) of chiral perturbation theory.  To determine if this
result is evidence against the presence of an IRFP for twelve flavors, two
questions must be resolved.  First, what is $F_\pi$ and is the condition
$L^{-1} \ll F_\pi$ satisfied on these configurations?  Second, what are the
expected eigenvalue distributions for twelve flavors if the low energy theory
is governed by an IRFP, or if $F_\pi L \lesssim 1$, and how do they compare to
the computed distributions?  We believe the technique of epsilon-regime
calculations can play an important role in future explorations of confining
field theories. So, it is essential to continue these calculations until the
apparent contradiction can be resolved.

\subsection{\label{sub:walking_Yang-Mills}Walking Yang-Mills}

In the previous section, the goal was to identify the minimum number of
flavors $N_f^\mathrm{crit}$ needed to impede confinement and spontaneous
chiral symmetry breaking in SU($N_c$) gauge theory, thereby allowing the
formation of an IR fixed point.  Just below this critical number of flavors,
it is plausible that Yang-Mills exhibits behavior at intermediate scales that
is more like the physics of the non-Abelian Coulomb phase than of QCD.  Of
course, at very long distances, the theory must eventually confine otherwise
it would be inside the conformal window.  Such behavior is called ``walking''
because the renormalized coupling is expected to run much slower or ``walk''
when compared to QCD \cite{Holdom:1981rm,Holdom:1984sk,Yamawaki:1985zg,
Appelquist:1986tr,Appelquist:1987fc}.

\begin{figure}
  \begin{center}
    \includegraphics[height=0.3\textheight]{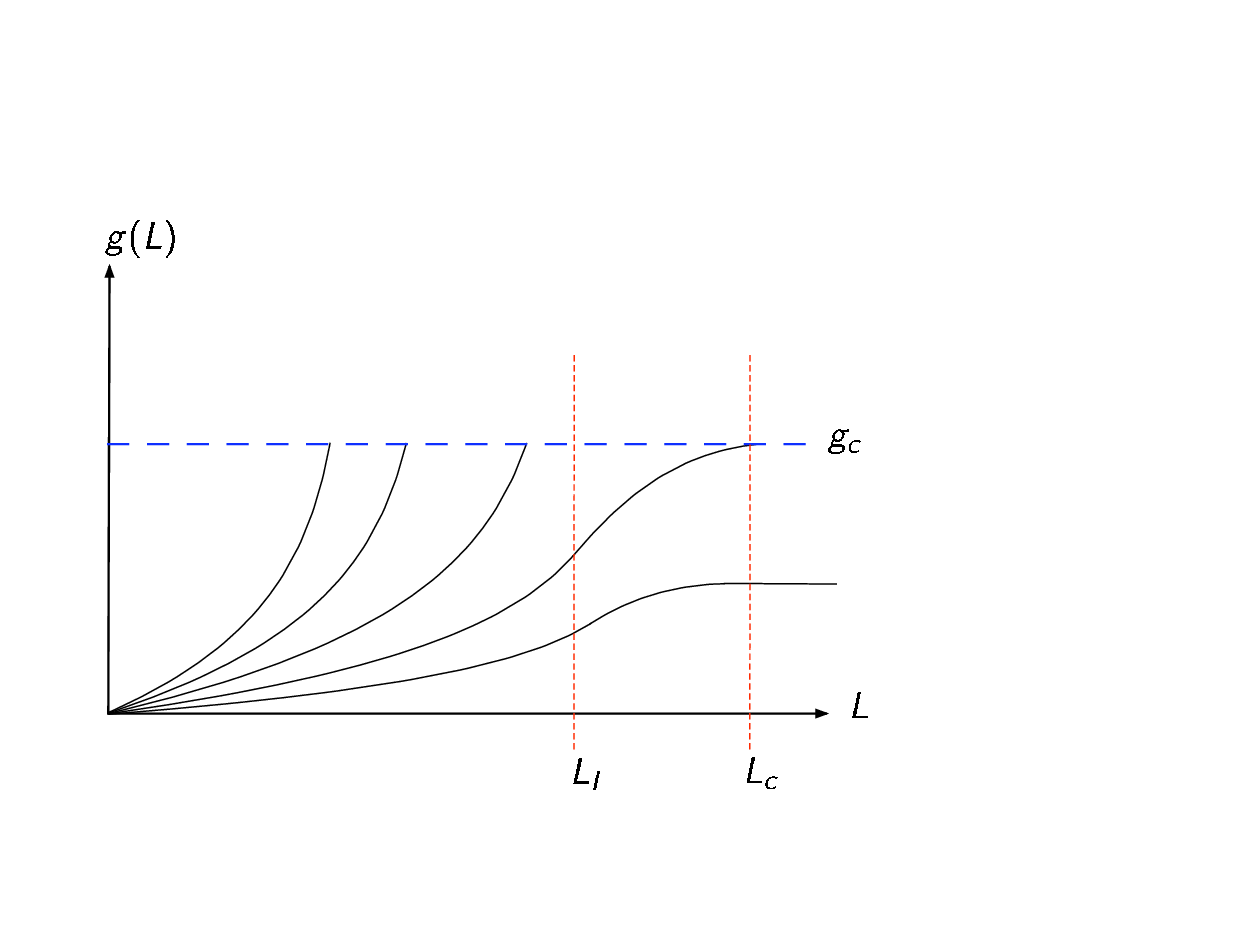}
  \end{center}
  \caption{\label{fig:walking}A cartoon of the dynamical scales of SU($N$)
    Yang-Mills.  $g(L)$ is a typical renormalized coupling. $g_c$ is the
    renormalized coupling required to trigger spontaneous chiral symmetry
    breaking (S$\chi$SB).  $L_c$ is the distance scale associated with 
    S$\chi$SB.  $L_I$ is the scale of the inflection point.  The curves are
    various $N_f$ increasing from left to right.  The rightmost curve is
    within the conformal window. A walking theory dynamically generates both
    scales $L_I$ and $L_c$.}
\end{figure}

The main feature of walking theories that is of interest to model builders is
that the pseudoscalar decay constant $f_\pi$ should be much smaller than the
chiral condensate $\left\langle q \overline{q} \right\rangle$, in sharp
contrast to QCD where the two quantities are of roughly the same order,
\textit{e.g.}\ $\left\langle q \overline{q} \right\rangle / f_\pi^2 \approx
25$.  In technicolor, for example, the scale of the Higgs VEV is set by the
pseudoscalar decay constant but the mass scale of technihadrons, including any
pseudo Nambu-Goldstone (NG) bosons are set by the much higher chiral
condensate.  If such a model were realized in Nature, no new resonances would
be observed (including the Higgs) below the lowest technivector meson
($\rho_T$) which is still too heavy to be observed directly at the Tevatron.
See the review by Hill and Simmons \cite{Hill:2002ap} for details.

Another potentially interesting feature of a walking Yang-Mills gauge theory
is that the spectrum of vector and axial-vector meson resonances may be
significantly different than QCD, and perhaps even inverted
\cite{Appelquist:1998xf}.  Such an inverted spectrum would give rise to a
negative $S$ parameter in contrast to the QCD-like behavior where $S$
increases with the number of flavors.  Thus, without detailed knowledge about
the low energy spectrum of a walking theory, precision electroweak constraints
do not apply because they assume a QCD-like spectrum.

\section{\label{sec:higher_reps}HIGHER FERMION REPRESENTATIONS IN SU($N$)
  YANG-MILLS GAUGE THEORY}

\begin{figure}
  \begin{center}
    \includegraphics[height=0.3\textheight]{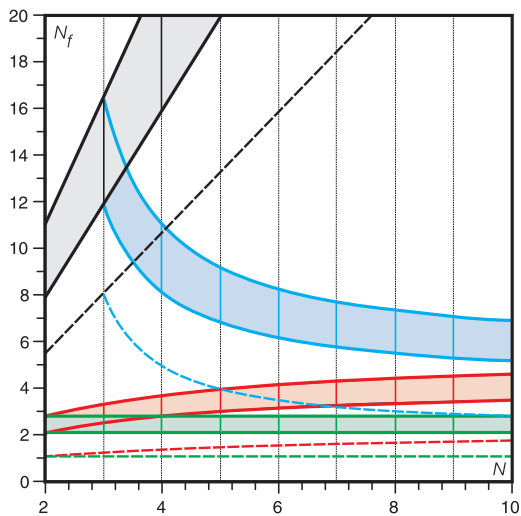}
  \end{center}
  \caption{\label{fig:PHRVA-D75-085018-Fig-1}Zero temperature phase diagram
    for SU($N$) Yang-Mills gauge theories with Dirac fermions in various color
    representations, taken from Ref.~\cite{Dietrich:2006cm}.  Representations
    from top to bottom are fundamental (grey), two-index antisymmetric (blue),
    two-index symmetric (red), and adjoint (green).  Shaded regions are
    estimates of the conformal windows.  Dashed curves indicate where the
    two-loop fixed point coupling diverges.}
\end{figure}

Recently, there has been a lot of interest in SU($N_c$) Yang-Mills theories
with $N_f$ fermions in higher representations of the gauge group.  As with
fermions in the fundamental representation, it is expected that the theories
confine if $N_f < N_f^\mathrm{c}$, are governed by an IRFP if $N_f^\mathrm{c}
< N_f < N_f^\mathrm{af}$ (the conformal window), and asymptotic freedom is
lost if $N_f > N_f^\mathrm{af}$.  Recent predictions of the extent of these
conformal windows, shown in Fig.~\ref{fig:PHRVA-D75-085018-Fig-1}
\cite{Dietrich:2006cm}, can now be tested by lattice methods.

There has been interest in the past \cite{Kogut:1985xa,Kogut:1986jt,
Karsch:1998wn,Karsch:1998qj,Lutgemeier:1998,Schulze:2004,Engels:2005te,
Engels:2005rr} in the SU(3) theory with two Dirac flavors in the adjoint
representation because of the possibility of two distinct finite temperature
phase transitions.  At higher temperatures, SU($2N_f$) chiral symmetry will
break to SO($2N_f$) but the global $Z(N_c)$ center symmetry of the gauge
action will remain unbroken until lower temperatures.  This theory is
apparently outside the conformal window although not much is known about the
zero temperature, low energy effective description.

Two years ago, there were some preliminary calculations done by Catterall and
Sannino \cite{Catterall:2007yx} studying SU(2) Yang-Mills with $N_f=2$ flavors
of adjoint fermions.  In the past year, they have extended their work with new
collaborators \cite{Catterall:2008qk}.  Two other groups have also presented
new calculations for this theory and presented their work at this meeting
\cite{DelDebbio:2008wb,DelDebbio:2008zf,DelDebbio:2008tv,Hietanen:2008vc,
Hietanen:2008mr}.  It is reassuring to see a great deal of consistency amongst
the groups as to the results of specific calculations.  However, larger volume
calculations are needed to bound the size of finite volume effects.

Two other groups also presented first results on SU(3) Yang-Mills with $N_f=2$
flavors of fermions in the two-index symmetric, or sextet, representation.
DeGrand and Svetitsky
\cite{Shamir:2008pb,Svetitsky:2008bw,DeGrand:2008dh,DeGrand:2008kx} presented
evidence using the Wilson fermion formulation that the low energy behavior of
the theory is governed by an IRFP on the basis of the running of the
renormalized coupling in the SF scheme, similar to the case of $N_f=12$
flavors of fermions in the fundamental representation described in
Sec.~\ref{sub:running_coupling}.  Their results are quite interesting.
Further calculations are done closer to the continuum limit would be welcome,
perhaps for $L/a=16$ as was typical for QCD calculations done by the ALPHA
collaboration.  N\'ogr\'adi presented preliminary results \cite{Fodor:2008hm}
for an epsilon-regime calculation of eigenvalue distributions for the same
theory using the overlap fermion formulation.  We believe the same comments
regarding the epsilon-regime calculation at twelve flavors apply here and look
forward to their results with dynamical fermions.

Much has been made of the fact that the conformal window is likely to extend
to much lower values of $N_f^c$ than is the case for the fundamental
representation and that this has implications for the value of the $S$
parameter and the possibility of walking behavior.  It is reasonable to expect
that the low energy dynamics of these theories will be quite different than
QCD and that naive estimates for the $S$ parameter are likely to be
unreliable, so we should wait for a direct calculation of the $S$ parameter in
these theories before drawing any conclusion about viability given the
precision EW constraints.

Finally, walking dynamics are useful for model builders because it produces
condensates $\left\langle\overline{Q}Q\right\rangle$ that are enhanced
relative to their natural scale $F_\pi$ as described in
Sec.~\ref{sub:walking_Yang-Mills}. This allows the breaking of ETC gauge
symmetry to be pushed to high scales to naturally suppress FCNCs while still
allowing for reasonable masses for standard model fermions.  We are unaware of
any ETC model involving technifermions in higher representations.  It would be
interesting to see to what extent the walking mechanism will be necessary
and/or useful in an ETC context should the LHC discover DEWSB by higher
representation fermions.

\section{ACKNOWLEDGMENTS}

We would like to thank the members of the Lattice Strong Dynamics (LSD)
collaboration for their many useful discussions and careful reading of these
proceedings: T.~Appelquist, R.~Babich, R.~C.~Brower, M.~Cheng, M.~A.~Clark,
M.~Kastoryano, T.~Luu, E.~T.~Neil, J.~C.~Osborn, C.~Rebbi, D.~Schaich,
R.~Soltz and P.~M.~Vranas.  We would also like to thank the organizers and
attendees of the \textsl{Workshop on Lattice Gauge Theory for LHC Physics},
May 2--3, 2008, Livermore, CA and the \textsl{Workshop on Dynamical
Electroweak Symmetry Breaking}, 9--13 September 2008, Odense, Denmark, where
there were many useful discussions that had impact on the remarks presented
here.


\appendix

\section{\label{app:finite_temperature}Flavor Dependence of Finite
  Temperature Transition}

Since the last comprehensive review of the flavor dependence of the finite
temperature transition (including $N_f \ge 4$) was two decades ago
\cite{Ukawa:1988ys}, it seems appropriate to revise and extend the summary
tables here for the bare critical coupling $\beta_c$ \textit{vs}.\ temporal
extent $N_t$, and where possible, the critical temperature $T_c$ in
dimensionful units.  Until recently, the preferred action for these studies
were the SU(3) Wilson gauge action and the unimproved Kogut-Susskind staggered
fermion action.  For $N_f$ not a multiple of four, some form of the rooting
trick was applied.

\begin{center}\begin{longtable}{c|c|c|c|c|c|c}
\caption{\label{tab:critical_temperature}Summary of critical temperature
calculations for SU(3) Yang-Mills with $N_f$=0--18 flavors, using the
unimproved Wilson gauge action and Kogut-Susskind staggered fermion action.}
\\

\hline\hline
$N_f$ & $N_t$ & $N_s$ & $a m$ & $\beta_c$ & $T_c$ (MeV) & refs \\
\hline\hline
\endfirsthead

\hline\hline
\multicolumn{7}{c}{Continued from previous page} \\
\hline\hline
$N_f$ & $N_t$ & $N_s$ & $a m$ & $\beta_c$ & $T_c$ (MeV) & refs \\
\hline\hline
\endhead

\hline\hline
\multicolumn{7}{c}{Continued on next page} \\
\hline\hline
\endfoot

\hline \hline
\endlastfoot

 0 &  4 & 16       & $\infty$ & 5.6908(2)   & 290.35(14) & \cite{Boyd:1996bx}    \\
 0 &  4 & $\infty$ & $\infty$ & 5.69254(24) & 291.54(16) & \cite{Iwasaki:1992ik} \\
 0 &  4 & $\infty$ & $\infty$ & 5.6925(2)   & 291.52(14) & \cite{Boyd:1996bx}    \\
\hline
 0 &  6 & 32       & $\infty$ & 5.8938(11)  & 293.47(59) & \cite{Boyd:1996bx}    \\
 0 &  6 & $\infty$ & $\infty$ & 5.89405(51) & 293.60(27) & \cite{Iwasaki:1992ik} \\
 0 &  6 & $\infty$ & $\infty$ & 5.8941(5)   & 293.63(27) & \cite{Boyd:1996bx}    \\
\hline
 0 &  8 & 32       & $\infty$ & 6.0609(9)   & 292.01(42) & \cite{Boyd:1996bx}    \\
 0 &  8 & $\infty$ & $\infty$ & $\approx 6.0625$
                                            & $\approx 292.75$
                                                         & \cite{Boyd:1996bx}    \\
\hline
 0 & 12 & 32       & $\infty$ & 6.3331(13)  & 290.94(52) & \cite{Boyd:1996bx}    \\
 0 & 12 & $\infty$ & $\infty$ & $\approx 6.3384$
                                            & $\approx 293.09$
                                                         & \cite{Boyd:1996bx}    \\
\hline\hline
 1 &  4 &  8 & 0.05  & 5.475(5)  &            & \cite{Fukugita:1987mb,
                                                     Fukugita:1988qs}  \\
 1 &  4 &  8 & 0.05  & 5.48      &            & \cite{Gavai:1988wd}     \\
 1 &  4 &  8 & 0.1   & 5.51(1)   &            & \cite{Fukugita:1987mb,
                                                     Fukugita:1988qs}  \\
 1 &  4 &  8 & 0.1   & 5.53(1)   &            & \cite{Irback:1988mm}    \\
 1 &  4 &  8 & 0.2   & 5.57(1)   &            & \cite{Fukugita:1987mb,
                                                     Fukugita:1988qs}  \\
 1 &  4 &  8 & 0.4   & 5.63(1)   &            & \cite{Fukugita:1987mb,
                                                     Fukugita:1988qs}  \\
\hline\hline
 2 &  4 & 16 & 0.01    & 5.265(5)   & 153.6(1.7) & \cite{Vaccarino:1991kn} \\
 2 &  4 &  8 & 0.0125  & 5.269(2)   & 153.3(7)   & \cite{Fukugita:1990dv}  \\
 2 &  4 & 12 & 0.0125  & 5.271(2)   & 154.0(7)   & \cite{Fukugita:1990dv}  \\
 2 &  4 &  8 & 0.025   & 5.286(2)   & 150.8(6)   & \cite{Fukugita:1990dv}  \\
 2 &  4 &  8 & 0.025   & 5.2875     & 151.3      & \cite{Gottlieb:1987eg,
                                                        Gottlieb:1988gr}  \\
 2 &  4 & 12 & 0.025   & 5.288(2)   & 151.4(6)   & \cite{Fukugita:1990dv}  \\
 2 &  4 & 16 & 0.025   & 5.291(1)   & 152.4(3)   & \cite{Vaccarino:1991kn} \\
 2 &  4 &  8 & 0.05    & 5.34(1)    & 151.5(3)   & \cite{Fukugita:1988qs}  \\
 2 &  4 &  8 & 0.05    & 5.32       & 146.5      & \cite{Gottlieb:1987eg,
                                                         Gottlieb:1988gr}  \\
 2 &  4 &  8 & 0.1     & 5.3825(25) & 138.8(4)   & \cite{Fukugita:1988qs}  \\
 2 &  4 &  8 & 0.1     & 5.375      & 137.7      & \cite{Gottlieb:1987eg,
                                                        Gottlieb:1988gr}  \\
 2 &  4 &  8 & 0.1     & 5.3825(50) & 138.8(7)   & \cite{Irback:1988mm}    \\
 2 &  4 & 12 & 0.1     & 5.376(3)   & 137.8(4)   & \cite{Irback:1988mm}    \\
\hline
 2 &  6 & 12 & 0.0125  & 5.415      & 144.1      & \cite{Bernard:1996cs}   \\
 2 &  6 & 12 & 0.025   & 5.445      & 140.8      & \cite{Bernard:1996cs}   \\
 2 &  6 & 12 & 0.025   & 5.4375     & 138.5      & \cite{Gottlieb:1988gr}  \\
 2 &  6 & 12 & 0.05    & 5.47       & 125.4      & \cite{Gottlieb:1988gr}  \\
 2 &  6 & 12 & 0.1     & 5.525      & 106.0      & \cite{Gottlieb:1988gr}  \\
\hline
 2 &  8 & 16 & 0.004   & 5.43--5.53 & 121--165   & \cite{Mawhinney:1992pc} \\
 2 &  8 & 16 & 0.00625 & 5.475--5.5 & 135--145   & \cite{Gottlieb:1996ae}  \\
 2 &  8 & 16 & 0.0125  & 5.54(2)    & 151(9)     & \cite{Gottlieb:1992ii}  \\
 \hline
 2 & 12 & 24 & 0.008   & 5.675(25)  & 168(15)    & \cite{Bernard:1996zw}   \\
 2 & 12 & 24 & 0.016   & 5.775(25)  & 186(13)    & \cite{Bernard:1996zw}   \\
\hline\hline
 3 & 4 & 16 & 0.01   & 5.105(1)   &            & \cite{Aoki:1998gia}     \\
 3 & 4 & 16 & 0.02   & 5.1235     &            & \cite{Liao:2001en}      \\
 3 & 4 &  8 & 0.025  & $\sim 5.1$ &            & \cite{Gavai:1987dk}     \\
 3 & 4 & 16 & 0.025  & 5.132      &            & \cite{Vaccarino:1991kn,
                                                       Aoki:1998gia,
                                                       Liao:2001en}      \\
 3 & 4 & 32 & 0.025  & 5.132      &            & \cite{Liao:2001en}      \\
 3 & 4 & 16 & 0.03   & 5.1396     &            & \cite{Aoki:1998gia,
                                                       Karsch:2001nf}    \\
 3 & 4 & 16 & 0.0325 & 5.1396     &            & \cite{Karsch:2001nf}    \\
 3 & 4 & 16 & 0.035  & 5.1505(5)  &            & \cite{Liao:2001en,
                                                       Karsch:2001nf}    \\
 3 & 4 & 16 & 0.04   & 5.1593     &            & \cite{Aoki:1998gia,
                                                       Karsch:2001nf}    \\
 3 & 4 &  8 & 0.1    & 5.25--5.3  &            & \cite{Gavai:1985nz}     \\
 3 & 4 &  8 & 0.1    & 5.30(5)    &            & \cite{Fucito:1984gt}    \\
 3 & 4 &  8 & 0.2    & 5.35(5)    &            & \cite{Fucito:1984gt}    \\
\hline\hline
 4 & 4 & 16 & 0.01   & 4.95        &            & \cite{Brown:1990by}    \\
 4 & 4 &  8 & 0.0125 & 4.919(6)    &            & \cite{Kogut:1987as}    \\
 4 & 4 &  8 & 0.025  & 4.94(1)     &            & \cite{Kogut:1987as}    \\
 4 & 4 &  8 & 0.025  & $\sim 5.03$ &            & \cite{Gavai:1987dk}    \\
 4 & 4 &  8 & 0.025  & 4.983(3)    &            & \cite{Fukugita:1990dv} \\
 4 & 4 & 16 & 0.025  & 4.99        &            & \cite{Brown:1990by}    \\
 4 & 4 &  8 & 0.0375 & 4.99(1)     &            & \cite{Kogut:1987as}    \\
 4 & 4 & 16 & 0.0375 & 5.02        &            & \cite{Brown:1990by}    \\
 4 & 4 & 16 & 0.05   & 5.04        &            & \cite{Brown:1990by}    \\
 4 & 4 &  8 & 0.1    & 5.05--5.15  &            & \cite{Polonyi:1984zt}  \\
\hline
 4 & 6 & 16 & 0.01   & 5.08        &            & \cite{Brown:1990by}    \\
 4 & 6 & 10 & 0.025  & 5.125(25)   &            & \cite{Kogut:1987as}    \\
 4 & 6 & 16 & 0.025  & 5.13        &            & \cite{Brown:1990by}    \\
\hline
 4 & 8 & 16 & 0.01   & 5.15(5)     & 93(15)     & \cite{Gavai:1990ny}    \\
\hline\hline
 8 &  4 & 16 & 0.015  & 4.58(1)     &            & \cite{Brown:1992fz} \\
\hline
 8 &  6 & 16 & 0.015  & 4.71(1)     &            & \cite{Brown:1992fz} \\
 8 &  8 & 16 & 0.015  & 4.73(1)     &            & \cite{Brown:1992fz} \\
 8 & 16 & 16 & 0.015  & 4.73(1)     &            & \cite{Brown:1992fz} \\
\hline\hline
10 & 4 &  8 & 0.1 & 4.70(10) &            & \cite{Fukugita:1987mb}  \\
10 & 4 &  8 & 0.2 & 4.85(5)  &            & \cite{Fukugita:1987mb}  \\
10 & 4 &  8 & 0.4 & 5.15(1)  &            & \cite{Fukugita:1987mb}  \\
10 & 4 &  8 & 0.6 & 5.35(5)  &            & \cite{Fukugita:1987mb}  \\
10 & 4 &  8 & 1.0 & 5.60(10) &            & \cite{Fukugita:1987mb}  \\
\hline\hline
12 & 4 &  8 & 0.1 & 4.50(10) &            & \cite{Fukugita:1987mb}  \\
\hline\hline
16 &  6 & 16 & 0.1 & 4.11--4.13 &            & \cite{Damgaard:1997ut}  \\
16 & 12 & 12 & 0.1 & 4.11--4.13 &            & \cite{Damgaard:1997ut}  \\
16 &  8 & 16 & 0.1 & 4.11--4.13 &            & \cite{Damgaard:1997ut}  \\
\hline\hline
18 & 4 &  8 & 0.1 & 4.25(15) &            & \cite{Fukugita:1987mb}  \\
\end{longtable}\end{center}

\begin{center}\begin{longtable}{c|c|c|c|c|c|c|c|c}
\caption{\label{tab:scale_setting}Summary of scale setting calculations for
SU(3) Yang-Mills with $N_f$=3--16 flavors, using the unimproved Wilson gauge
action and Kogut-Susskind staggered fermion action.  Empirical interpolating
functions are available for $N_f=0$ \cite{Edwards:1997xf} and $N_f=2$
\cite{Blum:1994zf}.} \\

\hline\hline
$N_f$ & $\beta$ & $m$ & $N_s$ & $N_t$ & $r_0/a$ & $a m_\rho$ & $a m_N$ & refs \\
\hline\hline
\endfirsthead

\hline\hline
\multicolumn{8}{c}{Continued from previous page} \\
\hline\hline
$N_f$ & $\beta$ & $m$ & $N_s$ & $N_t$ & $r_0/a$ & $a m_\rho$ & $a m_N$ & refs \\
\hline\hline
\endhead

\hline\hline
\multicolumn{8}{c}{Continued on next page} \\
\hline\hline
\endfoot

\hline \hline
\endlastfoot

3 & 5.115  & 0.015 &  8 & 32 & 2.10(30) &            &            & \cite{Nelson:2003tb,
                                                                          Nelson:2002si}    \\
3 & 5.1235 & 0.02  &  8 & 32 & 1.78(21) &            &            & \cite{Nelson:2003tb,
                                                                          Nelson:2002si}    \\
3 & 5.132  & 0.025 &  8 & 32 & 1.38(17) & 1.300(14)  &            & \cite{Liao:2001en,
                                                                          Nelson:2003tb,
                                                                          Nelson:2002si}    \\
3 & 5.1458 & 0.033 & 16 & 16 &          & 1.387(38)  &            & \cite{Karsch:2001nf}    \\
3 & 5.151  & 0.035 &  8 & 32 & 1.51(19) & 1.313(11)  &            & \cite{Liao:2001en,
                                                                          Nelson:2003tb,
                                                                          Nelson:2002si}    \\
3 & 5.3    & 0.01  & 16 & 32 & 3.9(1.0) &            &            & \cite{Nelson:2003tb,
                                                                          Nelson:2002si}    \\
\hline\hline
4 & 5.15   & 0.01  & 16 & 24 &          & 0.75(2)    & 1.10(6)    & \cite{Altmeyer:1992dd,
                                                                          Born:1993cq}      \\
\hline
4 & 5.2    & 0.01  & 12 & 24 &          & 0.96(5)    & 1.30(5)    & \cite{Born:1988jq,
                                                                          Born:1990ch}      \\
4 & 5.2    & 0.025 & 12 & 24 &          & 1.04(3)    & 1.58(4)    & \cite{Born:1988jq,
                                                                          Born:1990ch}      \\
4 & 5.2    & 0.05  & 12 & 24 &          & 1.12(2)    & 1.76(2)    & \cite{Born:1988jq,
                                                                          Born:1990ch}      \\
4 & 5.2    & 0.075 & 12 & 24 &          & 1.24(2)    & 1.90(2)    & \cite{Born:1988jq,
                                                                          Born:1990ch}      \\
\hline

4 & 5.35   & 0.01  & 12 & 24 &          & 0.695(35)  & 1.15(7)    & \cite{Born:1988jq,
                                                                          Laermann:1989zr}  \\
4 & 5.35   & 0.01  & 16 & 24 &          & 0.52(1)    & 0.77(3)    & \cite{Laermann:1989zr,
                                                                          Altmeyer:1992dd,
                                                                          Born:1993cq}      \\
4 & 5.35   & 0.025 & 12 & 24 &          & 0.83(2)    & 1.36(1)    & \cite{Born:1988jq,
                                                                          Born:1990ch}      \\
4 & 5.35   & 0.05  & 12 & 24 &          & 0.97(6)    & 1.60(1)    & \cite{Born:1988jq,
                                                                          Born:1990ch}      \\
4 & 5.35   & 0.075 & 12 & 24 &          & 1.12(2)    & 1.73(6)    & \cite{Born:1988jq,
                                                                          Born:1990ch}      \\
\hline
4 & 5.4    & 0.01  & 16 & 32 &          & 0.438(8)   & 0.690(21)  & \cite{Mawhinney:2000fw,
                                                                          Sui:2001rf}       \\
4 & 5.4    & 0.01  & 24 & 32 &          & 0.3742(44) & 0.5688(76) & \cite{Mawhinney:2000fw,
                                                                          Sui:2001rf}       \\
4 & 5.4    & 0.015 & 16 & 32 &          & 0.4736(59) & 0.7512(94) & \cite{Mawhinney:2000fw,
                                                                          Sui:2001rf}       \\
4 & 5.4    & 0.02  & 16 & 32 &          & 0.5008(28) & 0.7745(51) & \cite{Mawhinney:2000fw,
                                                                          Sui:2001rf}       \\
4 & 5.4    & 0.02  & 24 & 32 &          & 0.4872(26) & 0.7415(54) & \cite{Mawhinney:2000fw,
                                                                          Sui:2001rf}       \\
\hline\hline
8 & 4.65   & 0.015 & 16 & 32 &          & 0.522(7)   & 0.872(10)  & \cite{Brown:1992fz}     \\
8 & 5.0    & 0.015 & 16 & 32 &          & 0.484(7)   & 0.807(7)   & \cite{Brown:1992fz}     \\
\hline\hline
16 & 4.125  & 0.1   &    &    &          & 1.210(8)   & 1.90(1)    & \cite{Damgaard:1997ut} \\
16 & 4.25   & 0.1   &    &    &          & 1.123(6)   & 1.90(1)    & \cite{Damgaard:1997ut} \\
16 & 4.375  & 0.1   &    &    &          & 1.079(4)   & 1.737(3)   & \cite{Damgaard:1997ut} \\
\end{longtable}\end{center}

\bibliography{Fleming}
\bibliographystyle{apsrev}

\end{document}